\documentclass[acmsmall]{acmart}
\AtBeginDocument{%
  \providecommand\BibTeX{{%
    \normalfont B\kern-0.5em{\scshape i\kern-0.25em b}\kern-0.8em\TeX}}}

\setcopyright{acmcopyright}
\copyrightyear{2018}
\acmYear{2018}
\acmDOI{XXXXXXX.XXXXXXX}

\acmConference[Conference acronym 'XX]{Make sure to enter the correct
  conference title from your rights confirmation email}{June 03--05,
  2018}{Woodstock, NY}
\acmBooktitle{Woodstock '18: ACM Symposium on Neural Gaze Detection,
 June 03--05, 2018, Woodstock, NY} 
\acmPrice{15.00}
\acmISBN{978-1-4503-XXXX-X/18/06}

\usepackage{todonotes}
\usepackage{soul}
\usepackage{graphicx}

\begin{document}

\title[Examining ethics of LLM use in HCI research practices]{``I'm categorizing LLM as a productivity tool'': Examining ethics of LLM use in HCI research practices}




\author{Shivani Kapania}
\authornote{Both authors contributed equally to this research.}
\email{skapania@andrew.cmu.edu}
\author{Ruiyi Wang}
\authornotemark[1]
\email{ruiyiwan@andrew.cmu.edu}
\affiliation{%
  \institution{Carnegie Mellon University}
  \streetaddress{5000 Forbes Ave}
  \city{Pittsburgh}
  \state{PA}
  \country{USA}
  \postcode{15213}
}

\author{Toby Jia-Jun Li}
\affiliation{%
  \institution{University of Notre Dame}
  \city{Notre Dame}
  \country{USA}}
\email{toby.j.li@nd.edu}

\author{Tianshi Li}
\affiliation{%
  \institution{Northeastern University}
  \city{Boston}
  \country{USA}}
\email{tia.li@northeastern.edu}

\author{Hong Shen}
\affiliation{%
  \institution{Carnegie Mellon University}
  \city{Pittsburgh}
  \country{USA}}
\email{hongs@andrew.cmu.edu}

\renewcommand{\shortauthors}{Kapania and Wang, et al.}

\begin{abstract}

Large language models are increasingly applied in real-world scenarios, including research and education. These models, however, come with well-known ethical issues, which may manifest in unexpected ways in human-computer interaction research due to the extensive engagement with human subjects. This paper reports on research practices related to LLM use, drawing on 16 semi-structured interviews and a survey conducted with 50 HCI researchers. We discuss the ways in which LLMs are already being utilized throughout the entire HCI research pipeline, from ideation to system development and paper writing. While researchers described nuanced understandings of ethical issues, they were rarely or only partially able to identify and address those ethical concerns in their own projects. This lack of action and reliance on workarounds was explained through the perceived lack of control and distributed responsibility in the LLM supply chain, the conditional nature of engaging with ethics, and competing priorities. Finally, we reflect on the implications of our findings and present opportunities to shape emerging norms of engaging with large language models in HCI research.

\end{abstract}





\maketitle

\section{Introduction} 

\label{introduction}
The rapid development and adoption of large language models (LLMs) is reshaping the research and education landscape, sparking interest and concerns across many disciplines \cite{thirunavukarasu2023large, kasneci2023chatgpt, demszky2023using}.  
Large language models have also presented opportunities to complement research workflows in Human-Computer Interaction (HCI) research, including the analysis of qualitative data \cite{gebreegziabher2023patat, xiao2023supporting} and quantitative data \cite{mcnutt2023design, ning2023empirical}, replicating human-subject experiments in social sciences \cite{bail2023can}, and facilitating ways to simulate emerging social dynamics at scale \cite{park2023generative}.

However, in contrast to the excitement towards the potential of LLMs, a growing body of work has surfaced risks associated with these models \cite{weidinger2021ethical, kasneci2023chatgpt}, including misinformation, discrimination and exclusion, malicious use, and more. Cheng \textit{et al.} \cite{cheng2023compost} demonstrated how GPT-4 simulations of certain demographics (\textit{e.g.,} marginalized race/ethnicity
groups) and topics are highly susceptible to caricature. Participants' data privacy is also at risk, with evidence showing how LLM-based tools might leak sensitive information with or without malicious prompting \cite{zhang2023s, carlini2021extracting}.

Examining the use of LLMs holds particular relevance for HCI, given our frequent interactions with human subjects, engagement in community-collaborative efforts \cite{mackay1995ethics}, and interest in designing systems with a socio-technical approach \cite{bruckman2014research}. Indeed, the HCI community has demonstrated a long-standing commitment to understanding the impacts and ethical considerations of emerging technologies in research practices, dating back to the use of videotapes in the early 1990s \cite{mackay1995ethics}. Over the years, the SIGCHI research ethics committee has been facilitating open conversations about ethical challenges in our communities through research ethics town halls and panels at conference venues such as CHI \cite{fiesler2018research,munteanu2019sigchi,frauenberger2017research} and CSCW \cite{bruckman2017cscw,fiesler2021sigchi}. Researchers within the community have also organized several workshops and meetings to discuss the ethical challenges of HCI research, such as how to conduct large-scale user trials \cite{chalmers2011ethics}, how to engage with vulnerable populations \cite{liang2021embracing,antle2017ethics} and how to conduct research with public data \cite{fiesler2018participant}. 

Despite HCI's rich tradition of centering the discourse on research ethics, the rapid uptake of emerging LLM applications has brought renewed urgency to examine and collaboratively shape norms for LLM use \cite{shen2023shaping}. However, there is a gap in our understanding of HCI researchers' current practices surrounding LLMs, and uncovering them can offer a critical view into how they navigate ethical considerations. In this research, we ask: (1) How do HCI researchers integrate large language models in their projects? (2) What ethical concerns, if any, do they have regarding using LLMs? (3) How do HCI researchers approach and navigate those ethical concerns?

We report our results from 50 survey responses and 16 in-depth semi-structured interviews with HCI researchers using LLMs in their work. Across our participants, we find that LLMs were utilized throughout the entire HCI research process, from ideation to system development and paper writing. Large language models were perceived to open new possibilities for building tools and interactions, generating research ideas, and simplifying workflow for analysis and writing. We also came to see how researchers increasingly integrated LLMs into their everyday practice. 

Our participants anticipated a wide range of potential ethical issues associated with LLMs, such as potential harms in interacting with LLM outputs, privacy concerns, violation of intellectual integrity, and overtrust \& overreliance. 
While HCI researchers acknowledged these ethical considerations, in many cases, they were either unable to or only partially able to identify and address those ethical concerns in their own projects. Many participants highlighted their perceived lack of control with their position in the LLM supply chain \cite{widder2023dislocated}, a lack of established best practices, and competing priorities that took precedence over addressing ethical concerns.    

We begin by situating our study within prior work on LLM-based tools to support research workflows, scholarship on research ethics in HCI, and literature on the ethical issues with large language models. After presenting a detailed description of our methods, we present our findings on HCI research practices surrounding the ethics of LLM-based tools. Finally, we reflect on these results and present implications for the HCI research community. Taken together, our research underscores the importance of foregrounding research ethics if we are to continue integrating LLMs into our work practices. We call for engaging with IRBs and other regulatory institutions, redesigning effective informed consent processes, developing tools and processes to interrupt the LLM supply chain, providing learning opportunities for ethics of LLM use in HCI, and shifting existing academic incentive structures to foreground ethical considerations in research. 

\section{Related Work} \label{related_work}

\subsection{The Use of LLMs in HCI Research}
As the capabilities of generative AI in understanding context and generating natural language responses continue to advance ~\citep{openai2023gpt4}, large language models are being increasingly used by researchers as tools in Human-Computer Interaction (HCI) research.
Emerging work has explored the potential of leveraging LLMs to support the process of brainstorming and identifying research topics~\citep{systematic-review, ideas-are-dimes-a-dozen, suh2023structured}. Tools like CoQuest~\citep{coquest} have been developed for research question development through human-agent co-creation. In addition to research ideation, LLMs have been used as tools in human-AI co-creative design ideation~\citep{sketch-think-talk, ma2023conceptual} and writing support~\citep{zhang2023visar}.

Other members of the community have made efforts to develop LLM-powered applications for data generation and data analysis \cite{depaoli2023large}. Hämäläinen \textit{et al.} \cite{synthetic-hci-data} explored the potential of utilizing LLMs to produce synthetic user interview transcripts for piloting research ideas and designing interview protocols. Wei \textit{et al.} \cite{wei2023leveraging} created LLM-based chatbots for generating synthesized user self-reported data. Researchers have also developed LLM-based applications for qualitative analysis, including identifying themes and generating codebooks~\citep{dispensing, analysis-textual-data, gebreegziabher2023patat}. Prior work has also explored using LLMs for deductive coding~\citep{chew2023llmassisted} and for human-AI collaborative qualitative analysis~\citep{gao2023collabcoder, codebook-deductive, gebreegziabher2023patat}. In terms of quantitative analysis, tools such as Github Copilot\footnote{\url{https://github.com/features/copilot}} and GPT-4~\citep{openai2023gpt4} can enable researchers to transform natural language instructions into programming codes that assist quantitative data analysis and visualization. LLMs have been used as tools for quantitative data analysis in sampling, filtering, and analyzing survey data~\citep{JANSEN2023100020}, gaining insights from large corpus~\citep{oppenlaender2023mapping} and crowdsourcing social data~\citep{coding-crowdsourcing}. 

Researchers have also used LLMs as the underlying technology for system design and development.  
Lu \textit{et al.} \cite{lu2022bridging} and Petridis \textit{et al.} \cite{prompt-infuser} explored infusing prompt-based prototyping enabled by LLM into functional user interface design. Researchers also utilized LLMs to build task-oriented social simulations for LLM agents, aiming to study the intricate social dynamics of societal systems~\citep{gao2023s3, park2022social, li2023metaagents}. Other applications of integrating LLMs into the design and development of HCI systems have focused on leveraging LLMs for efficient prototyping~\citep{leiker2023prototyping, wu2022promptchainer} that can be potentially used by HCI researchers and designers. Our research extends this scholarship by empirically examining
practices and ethical challenges of integrating large language models into HCI research projects.

\subsection{Research Ethics in HCI}
The Human-Computer Interaction (HCI) community has long been engaged in discussions regarding ethics ~\citep{research-ethics-sigchi, research-ethics-past-five-years}, revolving around responsible conduct in human subjects studies~\citep{changes-in-research-ethics, bruckman2014research}, including privacy, informed consent, and institutional review boards (IRBs). These concerns are integral to ensuring ethical conduct and preventing harms to research participants~\citep{five-provocations}. Existing HCI guidelines of privacy focus on the protection of participants’ sensitive information, the anonymity of the participants, and the confidentiality of collected data~\citep{sep-ethics-internet-research, facebook-ethics}. The rights of autonomy and self-determination of participants require HCI researchers to establish transparent procedures for informed consent on the collection and analysis of personal information \cite{acm-codeofethics}. IRBs play a critical role in ensuring that HCI researchers meet their ethical obligations when conducting studies with human subjects ~\citep{research-ethics-irb, acm-humansubjectsresearch}. Moreover, prior scholarship underscores the need to mitigate potential biases in research design, data collection, and data analysis by refining existing ethical guidelines and protocols ~\citep{situational-ethics}.

The process of creating ethical guidelines for emerging technologies within the HCI community has drawn inspiration from ethical theories and professional standards. Mackay \cite{mackay1995ethics} highlighted the need to go beyond legal requirements and develop comprehensive guidelines for responsible behavior by learning from other disciplines. 
Vitak \textit{et al.} \cite{beyond-belmont-principles} encouraged interdisciplinary collaboration of building new ethics framework such as developing ethics heuristics for online data research to ensure responsible research. Despite the rich tradition of considering and iterating ethical guidelines in the HCI community, Computer Science (CS) researchers have demonstrated uncertainty about the applicability of ethical concerns pertaining to human subject studies to their research~\citep{cs-ethics-human-subject}, highlighting the necessity of educating CS and HCI researchers about research ethics.
Recent endeavors of integrating ethics in CS education such as adding systematic literature review on ethics in computing courses~\citep{integrate-ethics-in-ml, integrate-ethics-in-computing} have illustrated both the barriers and opportunities of educating CS researchers in ethics.

Recently, advances in AI have brought renewed urgency of research ethics to HCI and CS communities. Clark \textit{et al.} \cite{Clark2018} proposed approaches to assessing ethical risks of research involving digital data, informing the development of guidelines of ethical standards for research. Amershi \textit{et al.} \cite{guidelines-human-ai-interaction} provided guidelines for human-AI interaction design, emphasizing the importance of providing explanations of the systems and conveying consequences of user actions. Another line of research strives to bridge the gap between ethics and AI practices by establishing guidelines for safe, transparent, and trustworthy AI systems~\citep{bridge-gap}. With the rapid development of generative AI, Association of Computing Machinery (ACM) and the Association for Computational Linguistics (ACL) have established ongoing efforts to develop guidelines \cite{ACMPolicy, ACLPolicy}. The ACM policy on authorship requires the full disclosure of generative AI in the paper \cite{ACMPolicy}, while the ACL policy on AI-assisted tools on paper writing requires authors to elaborate on the scope and nature of their use \cite{ACLPolicy}. 

Even though underlying ethical guidelines may be broadly applicable, emerging technologies might present challenges to existing ethical review processes \cite{munteanu2019sigchi}. Our goal is not to create a taxonomy of all possible ethical concerns with the use of LLMs in HCI research; instead, we hope to extend the discourse on research ethics in HCI by documenting the ways in which researchers are responding to ethical considerations of LLMs in-situ and reflecting on potential ways forward. 

\subsection{Ethics of LLM use}
Extensive prior research has explored the ethical risks and harms related with language models~\citep{weidinger2021ethical, taxonomy-risks}.
The discrimination \& exclusion harms arise from the biased and unjust text in the training data of LLMs. Recent work identified the tendency of LLMs to display discrimination related to users' sensitive characteristics~\citep{xu2023llms} and demonstrated the gender and racial biases of LLM-generated content~\citep{fang2023bias}. The information hazards are the consequences of LLMs remembering private information in training data, posing the risk of privacy leaks~\citep{carlini2021extracting}. 

Privacy violation has been observed in LLM-based AI assistants where Personally Identifiable Information (PII) can be revealed by employing adversarial privacy-inducing prompts~\citep{lukas2023analyzing, mireshghallah2023llms}. 
The detection and mitigation of hallucinations in LLM generated text ~\citep{zhang2023sirens} remains a challenging problem ~\citep{chen2023llmgenerated}. This might lead to spread of misinformation, as LLM-based chatbots are often treated as fact-checking tools~\citep{zhang2023sirens}.

There is a growing concern about malicious activities arising from the generation of scams and phishing using LLMs~\citep{mozes2023use}. Recent research has also pointed out how the anthropomorphization of LLMs has the potential for manipulation and negative influence~\citep{deshpande2023anthropomorphization}. Finally, researchers have also documented the likelihood of LLMs increasing inequality and their negative effects on job quality, undermining creative economies, and more~\citep{taxonomy-risks}.

In response to these concerns, recent research has started to look into how to  assess~\citep{derczynski2023assessing} and mitigate~\citep{chen2023combating, mozes2023use} the harms of LLMs. For example, researchers are exploring privacy preserving strategies for language models in pre-processing, training, and post-training approaches~\citep{smith2023identifying}. For combating misinformation in LLMs~\citep{chen2023combating}, researchers have proposed several defense strategies, including integrating a misinformation detector~\citep{pan2023risk} and redesigning prompting methods to guide LLMs~\citep{zhang2023llmbased, pan2023risk}. Mozes \textit{et al.} \cite{mozes2023use} presented prevention measures for when LLMs were used for illicit purposes, arguing that red-teaming, safeguarding with RLHF and instruction following, and avoiding memorization and poisoning are potential solutions to the misuse of LLMs. Regarding human-computer interaction harms of LLMs, \citet{liao2023ai} provided a roadmap for improving the transparency and explainability of LLMs.

While recent efforts have begun to mitigate the broader ethical concerns of LLMs, the application of LLMs in HCI research presents unique ethical challenges. For instance, biases in the training data can adversely affect research practices, leading to issues with internal and external validity, reproducibility, efficiency, and the risk of proliferating low-quality research \cite{bail2023can}. Despite this, there is a notable gap in understanding how HCI researchers perceive and manage these ethical challenges in their day-to-day research activities. This study aims to fill this gap by exploring the unique ethical issues posed by the use of LLMs in HCI research and examining how researchers in this field are currently addressing these challenges.

\section{Methods} \label{methods}
In this research, we focus on examining the ways in which HCI researchers apply large language models (LLMs) across their research workflows and their ethical considerations for using LLM-based tools.
For a holistic view of LLM use practices, we employed a mixed-method approach with a sequential explanatory design \cite{ivankova2006using}. This involved conducting a survey with the goal of eliciting broad-brushed and higher-level perspectives from a wide audience. The survey was followed by semi-structured interviews to investigate, in more detail, the ways in which researchers approach the ethical considerations of using LLMs as part of their research activities. The qualitative data from the interviews was used to elaborate and explain the survey results (\textit{e.g.,} the rationale behind their approach to engaging with ethical concerns) and served as the foundation for our inquiry \cite{ivankova2006using}. Our research study was reviewed and approved by the IRB at our institution. We present our approaches to the survey and interviews in the following subsections. 

\subsection{Survey}

The goal of the survey was to identify the ways in which HCI researchers are using LLMs and any ethical considerations they have encountered in their projects. We conducted the survey using an online questionnaire implemented in Qualtrics and analyzed responses from 50 respondents.

\textbf{Participant recruitment.} We recruited survey participants through multiple channels: advertising on social media networks such as Twitter and LinkedIn, emailing direct contacts, and leveraging university distribution lists to which we had access. We began the survey by eliciting informed consent from respondents. No personally identifiable information was recorded about the respondents in accordance with our organization’s research privacy and ethics guidelines. The inclusion criteria for our survey were similar to the interviews, where we recruited researchers who are currently studying or working in areas related to Human-Computer Interaction (HCI) and have used large language models (LLMs) in their research. 

After the screening questions, we were left with n = 77 participants.  
Out of the 77 respondents, 50 completed all sections except for the demographics (which was optional). Among the 43 survey respondents who filled in the demographic questions, most reported working in academia (34),  industry (6), and non-profit (3) organizations. Researchers worked on projects within Human-AI Interaction (32), Design (13), Understanding People: Theory, Concepts, and Methods (12), Collaborative and Social Computing (10), and User Experience and Usability (10). In our sample, respondents were located in the United States (20), Afghanistan (5), Germany (3), Algeria (2), Hong Kong (4), China (1), Spain (1), Nigeria (1), Australia (1), Japan (1). On average, respondents had 4 years of experience working on HCI research projects.

\textbf{Questionnaire.} Our questionnaire consisted of 18 questions in total, with a mixture of multiple-choice (14) and open-text questions (4). We began the survey by describing LLMs as \textit{``a subset of generative language (and multimodal) models with increasing size as measured by the number of parameters and size of training data'' \cite{bender2021dangers} (}e.g.,\textit{ GPT-4, GPT-3.5, Llama 2, Vicuna, and more}). We then ask respondents to refer to their latest research project that involved the use of LLMs and respond to the following questions related to that project only. The survey was divided into three sections: (1) questions about their use of LLMs in their HCI research projects, (2) questions about how they engage with ethics of LLMs use in HCI research, and (3) demographic questions relevant to our study. In the first section, respondents were asked to describe the project in one sentence and share the primary research method, sub-area of HCI, and the stage of their research process where they incorporated LLMs in their project.

After understanding the HCI research project in which they used LLMs, we focused on their potential ethical considerations with the use of LLMs. We asked, \textit{``have [they] encountered or observed any ethical challenges related to LLMs in [their] research project?''}. This was followed by the question asking what the ethical challenges are in the form of a close-ended (with choices such as security and privacy, consent, harmful outputs, copyright issues, authorship, prefer not to say) and an open-ended question, respectively. We asked how they identified, potentially mitigated, and reported these ethical challenges. Finally, we also included demographic questions (optional to answer) about the respondent's type of institution, country, and years of experience in HCI.

\textbf{Analysis.} We computed a range of descriptive statistics using SPSS to better understand the approach of researchers to ethical concerns with LLMs. These included descriptive statistics to questions presented with multiple choice answers (\textit{e.g.,} the ethical challenges in using LLMs). In cases where questions were completed by a subset of respondents, we report question-specific response rates and the percentage of respondents who answered that question. Finally, we conducted a qualitative analysis of open-ended questions following the same approach to the interviews (see the following \autoref{sec:interview-methods} below). We performed multiple rounds of coding at the response level in conjunction with participants' survey ratings to surface high-level themes. We include direct quotes from our survey respondents in the Findings with the prefix `S\#' to differentiate them from our interview participants, which were prefixed with `P\#'.

\subsection{Interviews}
\label{sec:interview-methods}
Between October and November 2023, we conducted interviews with 16 HCI researchers who used LLMs in their research projects. Each interview had structured sub-sections beginning with the participants describing a recent project where they applied LLMs across any of their research activities to gain more context for the following questions. The responses in the survey informed the design of the interview questionnaire (e.g., further exploring the ethical concerns that were raised in the quantitative data). Participants could take a few minutes to look at their history with the LLM tools/applications. Our interviews focused on: (1) LLM use across the research workflow; (2) specific ethical considerations; (3) the process of navigating ethical considerations; (4) the role of IRBs; (5) the role of ethical frameworks and toolkits; (6) incentives and accountability. Each session focused on the researchers' practices and the associated ethical considerations. 

\textbf{Participant recruitment.} We recruited participants through a combination of distribution lists, professional networks, and personal contacts, using snowball and purposive sampling \cite{purposeful-sampling} iteratively until saturation. Our sample included researchers located in the United States (13), China (1), Singapore (1), and Germany (1). We interviewed 10 male and 6 female researchers. A majority of our participants work in academia (14), with two participants from industry. Refer to \autoref{tab:my_label} for details on participant domains, institution types, and demographics.

\begin{table}[t]
\centering
\small
\resizebox{\textwidth}{!}{%
    \begin{tabular}{l@{\hskip 0.3in}p{0.8\textwidth}}
    \toprule
    \textbf{Type} & \textbf{Count} \\
    \midrule
    \textbf{Institution Type} & Academia (14), Industry (2) \\\hline
    \textbf{Gender} & Female (6), Male (10) \\\hline
    \textbf{Expertise} & Human-AI Interaction (5), Privacy and Security (2), Computational Social Science (1), Educational Technology (1), Social Computing (1), Intelligent User Interface (1), UX Design (1), Creativity Tools (1), Mobile Computing (1), AI Ethics (1), Augmented and Virtual reality (1) \\\hline
    \textbf{Location} & USA (13), Singapore (1), China (1), Germany (1) \\\hline
    \textbf{Experience} & 0-5 years (5), 6-10 years (3), 10-15 years (1), N/A (7) \\
    \bottomrule
    \end{tabular}
}
\vspace{0.5em}
\caption{Summary of interview participants' institution types, domain expertise, and demographics.}
\vspace{-1em}
\label{tab:my_label}
\end{table}

\textbf{Interview moderation.} Given the geographical spread of our participants, the interviews were conducted online using video conferencing. We scheduled sessions based on the participants’ convenience and conducted all interviews in English. During recruitment, participants were informed of the purpose of the study. Informed written consent was obtained electronically for all interview participants. Participants were informed that they could refuse to answer any questions and/or ask for the recording to be paused at any time. Each session lasted about 40--60 minutes each.  We recorded interview notes through field notes and video recordings, which were transcribed verbatim for analysis. Each participant received a gift card of 30 USD for their participation.

\textbf{Analysis.} To analyze the qualitative data, we followed the reflexive thematic analysis approach by Braun and Clarke \cite{braun2006using, braun2019reflecting}. Reflexive thematic analysis foregrounds the researcher's role in knowledge production, with \textit{`themes actively created by the researcher at the intersection of data, analytic process, and subjectivity'} \cite{braun2019reflecting}. Two members of the research team independently read each interview transcript multiple times, starting with open coding instantiations of perceptions, ethical considerations, and challenges with using LLMs. They met regularly to discuss diverging interpretations or ambiguities. The entire research team met frequently to iteratively define themes based on our initial codes. In total, we transcribed 664 minutes of audio recording and obtained 336 first-level codes. As we generated themes from the codes, we also identified categories (units of analysis) together with a description and examples of each category.  These categories included (1) the use of LLMs across the research workflows, (2) ethical concerns, (3) approaches towards navigating ethical concerns, (4) barriers in addressing ethical issues, and (5) organizational structures. These categories were discussed and iteratively refined through meeting, diverging, and synthesizing into three top-level categories, presented in our Findings.  Since codes are our process, not product, IRR was not used \cite{mcdonald2019reliability, soden2024evaluating}. 

\section{Findings} \label{methods}
We begin by describing the ways in which HCI researchers are using large language models (LLMs) across various stages of their research process (\autoref{sec:research-stages-using-llm}). We then describe their ethical concerns (\autoref{sec:researchers-ethical-concerns}) and the four ways in which researchers engage with potential ethical concerns (\autoref{sec:mitigating-approaches}). In our interviews, the most commonly used LLM-based tools were ChatGPT, GPT-4 API, Bard and Bing Chat.

\subsection{Where do HCI researchers use LLMs in their everyday work?}
\label{sec:research-stages-using-llm}

\begin{figure}[t!]
    \includegraphics[width=0.9\textwidth]{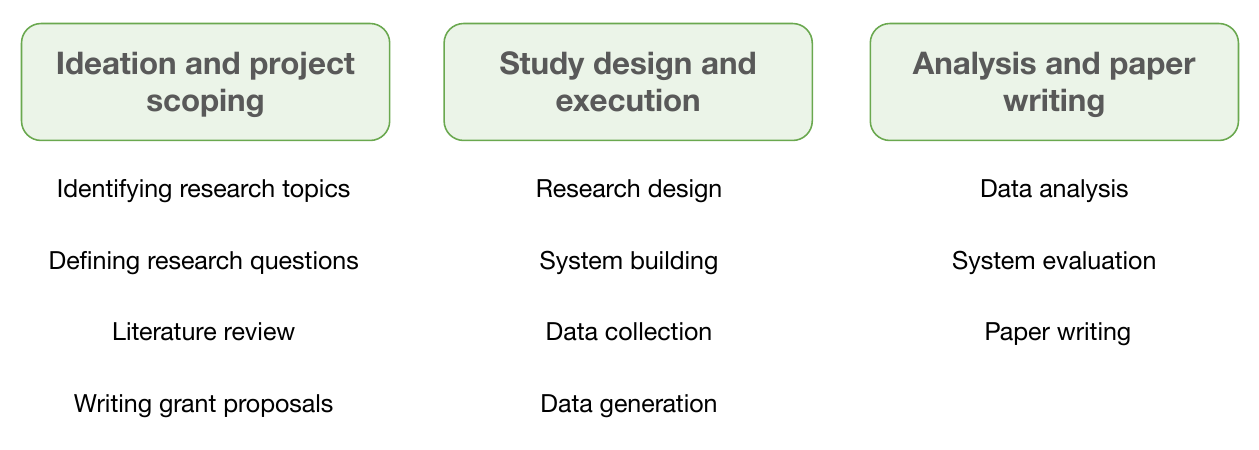}
    \caption{LLMs are used in various ways for ideation and project scoping, study design and execution, and analysis and paper writing. The figure illustrates a typical HCI study across our research participants. Not all HCI research projects would include all activities listed above (e.g., critical theoretical contributions).} 
    \vspace{-1.2em}
    \label{fig:label}
\end{figure}

HCI researchers referred to various parts of their research workflow where they integrated large language models, such as for ideation, literature review, study design, data analysis, system building and evaluation, and paper writing (\autoref{fig:label}). Overall, they perceived that LLMs open up new possibilities in their research, such that \textit{``if we can leverage LLMs the right way, it will enable us to do new cool things that will be genuinely empowering''} (P4). Our survey revealed that the stages where LLMs are most frequently used are paper writing (25) and study design (24), followed by project scoping (17) and system development (16), and then data generation (15), data collection (14) and data analysis (13). Below, we present the ways in which interview participants incorporate LLMs in their research practices.

Interview participants frequently used large language models in the \textbf{ideation and project scoping} phase, for tasks such as reviewing and synthesizing literature, discovering new research questions in their sub-field of HCI, and defining their research problems. P11 would input broad topic areas into the LLM to generate HCI research questions and subsequently refine them into concrete research objectives. Similarly, P10 would probe the LLM to \textit{``pretend that it is a career coach for [participant name]. What would [the LLM] recommend if [participant name] is writing their NSF career grants? This is a big thing for early career HCI academic researchers. What should [participant name] explore at the intersection of AI and cybersecurity?''} During brainstorming, HCI researchers found value in using large language models for a breadth-first search approach, enabling them to quickly generate a diverse range of ideas.

LLMs were also seen to have utility in \textbf{data generation, collection, and analysis}. Many participants mentioned how LLMs were especially productive in synthesizing information from web sources, that would otherwise require significant time and effort. P1 prompted the LLM to generate multiple arguments for and against hypothetical scenarios for use in a classroom setting. They noted how creating these research artifacts for data collection would have taken them weeks without the LLM. Additionally, HCI researchers integrated LLMs into their data analysis process, utilizing them for tasks such as for qualitative data coding, creating plots, and data visualizations. P7 applied LLMs for multiple aspects of open coding interview data, including (1) proposing new codes, (2) acting as a mediator between the coding team consisting of two members, and (3) generating the primary code groups.

HCI researchers described how they increasingly relied on LLMs in the \textbf{paper writing} stage. Participants shared their experiences of leveraging LLMs for iteratively refining their paper drafts, including searching for synonyms and fixing grammatical issues. LLMs offered a distinct perspective on aspects that the researcher \textit{``can hardly think about from [their] vantage point. Interacting with the LLM is like talking to other researchers and getting their feedback''} (P11). Researchers also mentioned using LLMs to generate paper reviews (P14), and provide suggestions for their writing content and style (\textit{e.g.,} as an alternative to Grammarly\footnote{\url{https://app.grammarly.com/}} (P4)). P14 spoke of using GPT-4 to \textit{``provide critical paper reviews, check grammar, the style, and the logic consistency''}. 

Finally, many HCI researchers used LLMs as a design material for \textbf{system development}, including system building and evaluation. Participants shared many use cases, such as developing new tools and interactions for library services, science communication, and AI-assisted writing. P4 expressed interest in exploring \textit{``what do large language models enable us to do in HCI with which our field has struggled for many decades.''} Finally, researchers in our study also discussed instances where they used large language models for software development, including day-to-day debugging and creating web interfaces with LLMs.

\subsection{What are HCI researchers' ethical concerns with the use of LLMs?}
\label{sec:researchers-ethical-concerns}

Researchers expressed a wide range of ethical concerns regarding the use of LLMs. Across our survey respondents, 30 reported observing ethical challenges associated with using LLMs, 10 expressed uncertainty, and 10 indicated no awareness of such concerns. Among the self-reported ethical challenges, the most common issues were data privacy (19) (including secondary use of participants’ data and data leak), authorship (16) (including disclosure of the use of LLMs in publications), harmful outputs (14), copyright issues (11) and consent (10). These concerns were more prevalent across research design and execution, as well as analysis and paper writing stages of the research (see \autoref{fig:label}). Below, we describe the specific ethical concerns highlighted by HCI researchers in our interviews.

\subsubsection{Harms of engaging with LLM outputs} 

Our interviews revealed a shared ethical concern of research subjects engaging with harmful outputs, especially if LLMs were integrated in systems and tools that directly interact with users. Over the last few years, scholars within HCI have increasingly shifted their efforts to working with and centering vulnerable populations (see \cite{cahill2007including, liang2021embracing}). P3 used large multimodal models to create a mental well-being and self-reflection tool and articulated their concern about LLMs generating uncontrolled outputs. LLM-generated content could disproportionately harm marginalized groups through exposure to socially harmful biases and stereotyping behaviors. Hate speech and exclusionary and discriminatory language could cause psychological and representational harm to research participants. P16, developing LLM-based roleplay robots, articulated how \textit{``[LLM] may exaggerate or diminish some of the capacity with which the robot has already been equipped. It may introduce another layer of bias [toward people with disabilities]. ''} In addition to the concerns about biased LLM outputs, P14 also worried about the possibility of LLM facilitating the \textit{``surveillance and monitoring of intimate partners''} by providing links to various spy tools. 

Many researchers expressed concerns about large language models impacting their research pipeline by generating seemingly authoritative but fabricated information. LLM hallucinations during paper writing could mislead authors and if published, would undermine trust in knowledge production processes. P10 used LLMs for literature review and worried that \textit{``ChatGPT might make up titles of things that simply do not exist.''} Participants highlighted the need for vigilance in identifying these hallucinations, especially when LLMs produced fake citations or mismatched paper references. Beyond paper writing, researchers indicated the possibility of inheriting biases from LLMs in ideation, research design, and evaluation. P2 articulated how LLMs could introduce \textit{``gender bias in the entire computational pipeline-- in places that we've never had gender bias before because we've never used LLMs before, and because we used to come up with our own solutions''}

Participants pointed out the tendency of LLMs towards generating homogenous content, where the model would prioritize generalization and converge diverse perspectives into standardized outputs. P8, working on sensemaking tools for product selection using LLMs, described how they \textit{``want the selection criteria to encompass a diverse set of opinions rather than just focus on the perspective of a single demographic.''} This tendency of LLMs towards \textit{``flattening human diversity and nuance''} (P9) raised concerns among researchers who emphasized the importance of capturing the complexity of lived experiences within the context of their research. Participants also noted concerns that using LLMs for editing their paper drafts could influence researchers' paper writing styles, streamlining towards homogeneous and \textit{``bland way of writing''} (P9). Participants also discussed how LLMs' training data, often reflecting Western morals and values, could contribute to this homogeneity.

\subsubsection{Threats to privacy of participant data} 

Researchers in our study expressed anxiety about how data input into LLMs is being used by LLM providers and the potential for violations of privacy. They worried about breaches of private or sensitive information involving various forms of participant data, such as transcripts, audio recordings, and application-specific log data. P7 used LLMs for qualitative data analysis and discussed how \textit{``privacy issue is the main concern for us because in many cases, the audio transcription is not supposed to be put into ChatGPT''.} P16 also shared the concern that users' navigation data is extremely sensitive and could lead to material physical harm if uploaded to LLMs.

Many participants emphasized their concern of confidential and personally identifiable information leakage as a result of sharing user data with LLM providers. P8 mentioned the risk of confidential information being incorporated into LLM training corpus and the potential for security leaks, where \textit{``backend messes up and user's private information shows up in someone else's chat history''}. Relatedly, some participants expressed concerns about the possibility of unconsented data, for example, \textit{``explicit sexual content''} (P2), being a part of the training corpus for large multimodal models.  While some believed that manually erasing sensitive information could mitigate these concerns, others relied on the efficacy of APIs in safeguarding user data. 

\subsubsection{Violations of intellectual integrity}

Interview participants raised ethical concerns about the intellectual integrity of using LLMs in HCI research. A central theme of these concerns was the ambiguity of ownership of LLM-generated text and visuals. Many participants who were co-creating with LLMs also highlighted the difficulty in attributing what portion of the content is the researcher's original work and what is generated by the LLMs when they are refining and co-creating with the LLM outputs. Participants discussed the ambiguity related to plagiarism when LLM outputs are part of the research contribution. For example, interviewee P10 would consider crediting the LLM \textit{``where a substantial amount ends up in a publication''}. While most participants questioned the extent to which one can claim ownership of LLM-generated content, especially in the paper writing stage, some believed that \textit{``personhood is pretty important for attribution''} (P2). 

In addition, HCI researchers also raised concerns about the reproducibility of the research results obtained via LLMs, highlighting the potential for an illusion of efficacy if LLMs work well in some cases but fail to generalize to others. P2 described using LLMs in HCI research as a \textit{``computational Wizard of Oz''} method and continued on to point out how the quick and opaque updates of LLMs are another barrier to reproducibility, noting that researchers \textit{``don't really have control of what version of GPT they are talking to and something that might work in a previous version won't work so much in future versions''}.

\subsubsection{Overtrust and overreliance on LLMs}

Participants also noted ethical considerations of overtrust and overreliance on LLMs. They expressed concerns that research participants who were directly interacting with LLMs and were unaware of their biases could overestimate LLM capabilities and place unwarranted trust in their suggestions. P13, who built LLM-powered creativity support tools, was concerned that \textit{``[research participants] may be misled by the content, even if [the content] contains wrong information or wrong references.''} 
They also expressed concern about researchers’ overreliance on LLMs, which could compromise the quality and creativity of HCI research. P3, using LLMs for creative ideation for their research, warned about the risk that researchers might uncritically accept the information provided by LLMs as factual and integrate them into their research plans. Participants mentioned how the overuse of LLMs among HCI researchers could impede creativity (\textit{e.g.,} \textit{``LLMs are useful but not creative''} (P15)) and raise questions about the long-term impact of such overreliance on LLMs on trust within the academic community.

\subsubsection{Environmental and societal impacts}
Interview participants identified a range of other higher-level ethical considerations that extended beyond their specific research project or pipeline. Some researchers were concerned about the environmental degradation due to the extensive electricity usage and hardware deployment of building \textit{larger} LLMs. For instance, P2 mentioned \textit{``we have all these models that are competing against each other, that's like millions of dollars of electricity and computer components, [which is] really bad for the environment''}. Participants articulated anxieties that the wide adoption of LLMs could lead to inequitable distribution of benefits-- \textit{``we certainly have not addressed the social issues. I think a lot of people will, if not losing their jobs, be substantially diminished in terms of their utility to their workplace''.}

\subsection{How do HCI researchers approach the ethical concerns of LLMs?}
\label{sec:mitigating-approaches}

In our study, we focused on understanding how HCI researchers are currently navigating ethical considerations about the use of LLMs in their projects. For the majority of HCI researchers in our interview study, the ethical concerns they discussed (\autoref{sec:researchers-ethical-concerns}) remained largely speculative. Although our participants reported their awareness of a diverse set of potential ethical concerns around the use of LLMs, they were either unable or only partially able to identify or address those ethical concerns in their own projects. Researchers described their current strategies towards ethical concerns-- often manifesting as forms of \textit{inactions or workarounds}-- such as engaging LLM ethics in a conditional and reactive manner, inconsistent disclosure practices, limiting their LLM use and reflecting on their co-creation process, as well as delaying responsibility in research accountability. In most cases, these strategies did not directly address the underlying ethical challenges; instead, they served as temporary fixes to manage immediate concerns. Below, we capture these varied strategies to LLM ethics and return to their implications in section 5.

\subsubsection{Conditional and reactive engagement with LLM ethics}

HCI researchers often emphasized that the need to and approach towards addressing ethical concerns was conditional on various factors such as the use casse or domains of study. Many HCI researchers invoked the specifics of their research domain to justify why they did not need to foreground ethical concerns. For example, when participants categorized their research as low-stakes, they spoke of how commonly anticipated ethical issues associated with LLMs were not applicable to their work, and they did not need to take proactive measures. For instance, P11, who focused on creativity support in writing, considered their research non-sensitive and did not find it necessary to intervene to prevent unsafe generated text. Some participants believed that the potential harms caused by LLMs were comparable to those of social media. Therefore, they felt that engaging in ethical considerations was unnecessary if LLMs were not being used in high-stakes domains, such as providing medical advice:  
\textit{``typical LLM like ChatGPT was created to support people browsing the internet. People already see toxic content online, right? So we always claim this tool is no more harmful than typical social media''} (P13). If researchers explored topics deemed safe or if users did not directly prompt the model, there was a perception that the LLM was unlikely to generate unsafe content\footnote{In contrast, prior work has demonstrated that LMs can generate unsafe content from seemingly innocuous prompts \cite{gehman2020realtoxicityprompts}.}.

Such `conditional engagement' sometimes resulted in a more reactive, rather than proactive, approach to ethical considerations. For example, when deliberating whether LLMs should be attributed and held accountable for model-generated content, P2 emphasized \textit{``I think many people are very hasty to say yes or no. And I think that's not the answer. The answer is always in a gray area.''} They continued to emphasize this ``wait and see'' approach, discussing the potential benefits of model hallucinations and highlighting their role in promoting divergent thinking by introducing specificity that the user may not have considered. Some participants mentioned being aware of frameworks and toolkits (such as Perspective API \cite{PerspectiveAPI}) designed to address ethical issues, but they never incorporated these tools into their own research processes. P3 justified this by discussing how a significant portion of their HCI studies were conducted in a laboratory setting. The ethical concern related to participants encountering harmful outputs generated by LLMs was less probable in a short usability test, while serious issues could occur in longitudinal studies when the participants heavily use the system.

\subsubsection{Limited disclosure practices}

In our study, HCI researchers positioned large language models as everyday tools within their research practice. As a result, participants did not believe it was necessary to formally report their usage of LLMs to study
participants, the Institutional Review Board (IRB), and/or the broader academic community. In particular, participants described a shift towards the tacit incorporation of LLMs: they slowly transitioned from research tools that were a part of their regular practice to using LLMs. This change was partly due to the perception that LLMs are \textit{`fancier'} (P11) and more advanced versions of previously used tools. P10 explained, \textit{``I advise my students that they are allowed to use any generative AI tool just like they would use other productivity tools. So I'm categorizing LLMs as a productivity tool.''}

When using LLMs in paper writing, participants drew a parallel between LLM reporting practices and their approach to previous tools. They emphasized that if, for example, tools like \textit{``grammar assistance by Grammarly, word check by Google Docs, or accessibility checking by Acrobat pro''} (P10) were not explicitly reported, the same should also hold true for large language models. Researchers expressed reservations, suggesting that reporting of LLM-based tools might call into question the validity of their work. P10 captured this perspective: 

\begin{quote}
    \textit{I would not feel it is appropriate to say Bing Chat was used to define the initial structure of X and Y sections of the paper. To me, it is not gonna be helpful in researchers assessing the credibility or validity of the work. It is just like a meta issue about how the actual document was formed and refined. And what really matters in that case is the output. Is it easy to read? Did you find it useful? You know, that's what I care about. So, in those cases, I do not credit the LLM.}
\end{quote}

The complexity of describing LLMs to audiences without technical background further impacted participants' willingness to disclose the specific uses of large language models in research. In some cases, researchers chose to characterize their LLM use simply as \textit{`AI models'} to their research participants so as not to \textit{``confuse them with what is a language model.''} The rationale behind this approach was the perception that the broader public (target population in this case) tend to view AI in a homogeneous manner, and \textit{``to them, there isn't much difference between how different AI systems work, or which is a large language model, right?''} (P8). Researchers also justified this approach by highlighting their intention not to burden participants with the need to modify their behavior and \textit{`` think about how to interact with an LLM''} (P15). Finally, participants also noted cases where explicit disclosure about LLMs was rather unnecessary if research participants or system users are not directly exposed to LLMs. For example, if LLMs were used for ideation or analysis, participants felt that explicit communication about their use was not imperative.  

\subsubsection{Restricting LLM use and reflecting on the co-creation process}

A recurring sentiment among the researchers in our study was a perceived lack of control in addressing ethical concerns related to large language models. As a result, they developed various workarounds, such as restricting the use of LLMs to a limited set of tasks,  avoiding directly integrating LLM-generated inputs into their work, and hosting group reflection sessions. The limited visibility and decision-making capability were particularly evident in issues of privacy and data leaks, where the reliance on LLMs provided by large companies diminished individual control. Simultaneously, participants expressed a mistrust towards LLM providers. For instance, P12 discussed how LLM providers’ claim to protect data privacy and not using it for their training \textit{``is very problematic because such claims don't really articulate what they mean by not using the data. How can an external researcher validate this claim?''} Participants' mistrust towards LLM providers was exacerbated by the lack of clarity and transparency in data usage policies.

To navigate the \textit{`unknown territory'} of LLMs, where HCI researchers were not fully aware of the capabilities and risks, they would often restrict their usage of LLMs to a limited set of tasks. Most HCI researchers avoided directly integrating LLM-generated outputs into their work. Rather than accepting the initial output without scrutiny, they would iteratively refine and carefully verify before incorporating it into their research artifacts. In the context of paper writing, researchers would ensure the text aligns with the draft they had input into the system. Many participants elaborated on their practice of co-creating with LLMs as a precautionary measure. This would involve relying on their personal judgment on how much to use the LLM suggestions. This cautious and iterative co-creation process served as a strategy in the absence of clear usage norms. 

Indeed, many participants shared the view that LLMs are still evolving, and there is a lack of comprehensive guidelines on navigating ethical considerations. P9, primarily using LLMs for paper writing, mentioned how \textit{``LLMs are still an unknown territory, so people don't know how to react, I assume.''} Our participants expressed discomfort with using any new LLM-generated content in their papers. For example, P15 was hesitant to use the LLM as anything more than a spell checker. Researchers highlighted the importance of applying LLMs in a way that is consistent with their practices with previously used tools to mitigate potential unintended consequences. P9 spoke of their experience as a reviewer for major HCI conferences like CHI, where they did not encounter any guidelines regarding the disclosure of using ChatGPT for writing. Finally, they went on to emphasize that the cost and accountability of using LLMs was still \textit{``ad-hoc and unclear.''}

In cases where participants recognized the possibility of ethical concerns with LLMs, they often struggled to identify specific issues to address. As P16 pointed out: \textit{``the main problem is that I don't know what bias it has, and I don't know how to figure it out.''}  P4 discussed how it is crucial to \textit{``not pretend that this ethical consideration does not exist, which some people do. We are trained in human-computer interaction, and so we are well equipped to reason about it or at least understand that these concerns exist, but obviously addressing it is very difficult.''} In many cases, when people did not have mechanisms to identify or address ethical concerns, they felt it was important to at least acknowledge them and spread awareness about these issues. 

To account for the uncertainties with the responsible use of LLMs, some researchers chose to host group reflection sessions. 
This involved engaging in discussions with collaborators or advisors to determine the appropriate approach for navigating any ethical considerations. P10 exemplified this approach by holding regular team meetings to address any questions or concerns related to using LLM-based tools or methods. After an internal discussion with their team, P9 too, decided not to use LLMs for qualitative analysis. A collaborative decision-making process was a common practice to determine where it is appropriate to use LLMs.

\subsubsection{Delaying responsibility in research accountability}
Finally, our participants indicated that determining who bears responsibility for the ethical implications of using LLMs is challenging. At times, they expressed reluctance towards stringent regulatory guidelines and opted to postpone dealing with ethical issues in their projects.

In discussions about accountability with the use of LLMs within HCI research, participants noted the ambiguity in determining where responsibility lies regarding the ethical concerns of using LLMs. They highlighted the notion of \textit{distributed responsibility} across the AI supply chain, emphasizing that multiple stakeholders share the responsibility for ensuring that the use of LLMs within research is ethical. Participants highlighted the LLM provider's responsibility to implement safeguards preventing users from sending sensitive personal data to the model. For instance, P8 discussed their confidence in LLM providers, sharing how they \textit{``believe OpenAI has done a decent job in terms of making a model safe in general.''} 
P2 presented a similar perspective: 

\begin{quote}
   \textit{``When professionally something wrong happens, it needs to follow the chain of command. So, who is the person most directly responsible? It's actually the user, right? The user did something with my system, and it created a harmful output, and then the user will say, well, the system wasn't designed well enough for me to know it's gonna create this harmful output. I should have been warned. In that case, the responsibility falls on me. And then I could move it up the chain of command to say my advisor shouldn't have let me release this system if it was gonna produce harmful output or I could say OpenAI is irresponsible for releasing a project that they are publicizing as being broadly accessible and safe.''} 
\end{quote}

Partially due to the difficulty in determining ``distributed responsibility'' for some HCI researchers, addressing ethical concerns could be relegated to future work. They viewed it as imperative, as HCI researchers focused on system-building, to explore emerging technologies and develop new tools and interactions with them. As articulated by P6, \textit{``we are the system builders, and we think about what are the unique benefits that the new technology can bring to the users and enhance the user capabilities.''} This perspective underscored the desire to continue building tools, even as researchers acknowledged the biases associated with LLMs.

Sometimes, participants even expressed a resistance towards prescriptive regulations, which they perceived would suppress innovation in HCI research. 
They emphasized that external constraints on LLM usage, especially given its role in everyday research practices, would slow down their research. LLMs were also perceived to lack the stability necessary for regulation. As P12 described, \textit{``the challenge is the regulation needs to come much later rather than early. If you come up with regulations too early, you may kill innovation and on the other hand, you don't know what you want to regulate because the representation of the product hasn't been stabilized yet.''}

\section{Discussion
} \label{discussion}
Our results highlight the various ways in which HCI researchers integrated LLMs in their research projects,  for the purpose of research ideation (\textit{e.g.,} finding novel research areas), data collection and analysis (\textit{e.g.,} qualitative coding), preparing artifacts (\textit{e.g.,} drafting research publications), and more. Many participants were aware of the potential ethical concerns related to the use LLMs in HCI research, including the harms of engaging with LLM outputs, threats to the privacy of participant data, intellectual integrity, and environmental and societal impacts. However, the sociotechnical assemblage of LLMs mediated researchers' diverse approaches to these ethical concerns. 

While HCI researchers are increasingly using LLM-based tools within their `everyday' practices, 
our research community still lacks well-established guidelines and best practices, contributing to the complexity of navigating ethical considerations. In addition, researchers noted a perceived lack of control over the functionality and outputs of LLMs. Unlike previous tools where researchers had more predictability and influence over the behavior, LLMs presented outputs that may not always align with explicit instructions. Finally, many participants spoke of the challenges of navigating ethical concerns within the broader ecosystem of the ‘LLM supply chain’ \cite{cobbe2023understanding}.
Below, we present the implications of researchers’ approaches and opportunities for HCI researchers to better engage with ethical considerations of LLMs as part of their projects, with a goal to support the formation of emerging ethical norms in LLM-impacted HCI research. 

\textbf{Proactively engaging with IRB and other regulatory institutions.} Within the U.S. context, Institutional Review Boards (IRBs) are responsible for reviewing and monitoring research activities to protect the rights and welfare of human research subjects \cite{IRB}. One of the primary responsibilities of the IRB office is to assess whether ``risks to subjects are reasonable in relation to anticipated benefits’’ \cite{IRBCommonRule}. Given their expertise in ensuring ethical conduct in research practices, the IRB office may be well-positioned to advise on identifying ethical concerns and potential prevention strategies. To establish whether a research study will undergo a comprehensive review, they determine if it is a minimal risk study where the ``probability and magnitude of harm anticipated in the research are not greater than those ordinarily encountered in daily life’’ \cite{IRBCommonRule}.

Our findings reveal that most HCI researchers did not consider it necessary to report their usage of LLMs to the IRB, in part due to their perception of LLMs as everyday tools. According to some, their scope of use did not justify including additional details, while others did not anticipate their use of LLMs at the time of submitting the IRB application. For those who disclosed their use of LLMs to the IRB, they used the minimal risk rule \cite{IRBCommonRule} to indicate that research subjects would typically encounter harmful outputs in their daily life as well. However, these practices can have short-term and long-term adverse implications. In the short term, the opacity in research practices (\textit{e.g.,} which tools are used to generate research artifacts or analyze data) might lead to challenges in replicating studies \cite{TransparentStatsJun2019} and understanding the motivation and effects of methodological decisions \cite{changes-in-research-ethics}. In the long term, the consequences of limited disclosure might extend beyond individual studies to shaping the trajectory of the community, for example, by informing which research topics we pursue. 

We invite HCI researchers to proactively engage with the IRB at the time of study design to unpack the likelihood and ways in which any potential LLM use might harm our participants. This includes careful reflection and documentation of any implicit or anticipated use during the project planning stage. Furthermore, researchers should actively implement mechanisms to monitor LLM outputs and any adverse impacts on participants throughout the project life-cycle, especially if used for system building. In addition, openly communicating and collaborating with IRB members to uncover potential harms is also important to increase transparency. 
Embracing guidelines, policies and oversight is essential for ensuring responsible development and use of these technologies. \citet{Bockting2023} urged scientists to oversee the use of generative AI because \textit{``controlling developments in AI will require a continuous process that balances expertise and independence.''} 
We call for collaborative efforts between researchers, policymakers, and generative AI companies to create a set of `living guidelines' for the responsible use of generative AI in research~\citep{Bockting2023}.

\textbf{Re-examining the informed consent process.} In our study, researchers acknowledged the limited transparency to their study participants, either by not disclosing LLM use (\textit{e.g.,} considering it irrelevant unless directly involved in system building), or by characterizing the underlying technology as Artificial Intelligence (AI). This was often rationalized as a way to avoid overwhelming participants with seemingly unnecessary technical details. However, terminologies can be powerful too: they shape narratives around capabilities and set expectations for use \cite{cave2018portrayals}. AI can appear \textit{`magical’}, where systems are considered reliable and free of bias \cite{kapania2022because}. In contrast, LLMs have specific evidence-based, known risks \cite{taxonomy-risks} that must be clearly communicated to research subjects. For example, in March 2023, a bug allowed some ChatGPT users to see the titles of other users’ chat histories \cite{chatgptleak}. It is important to recognize that our research practices can place participants’ safety at risk \cite{shklovski2012googling}. 

Researchers should carefully attend to the informed consent process for projects that may involve the use of LLMs across different research stages. There is a well-established body of work on informed consent in research that advocates for moving beyond obtaining consent as `instrumental in nature' \cite{hamilton2009ethics} only to satisfy regulatory and reporting obligations \cite{klykken2022implementing}. Research participants need to have a clear understanding of the objectives of the research, methods, procedures, as well any foreseeable risks \cite{InformedConsent}. How might informed consent facilitate a collective, continuous sense-making process between the researcher and participant to understand the implications and ethical concerns of using LLMs? Nonetheless, it is crucial to highlight that documentation, transparency, and effective informed consent might help identify and prevent issues, but they cannot substitute for efforts to mitigate ethical concerns with the use of LLMs.

\textbf{Developing tools, methods and processes to interrupt the LLM supply chain. } Over the last few years, progress in Natural Language Processing (NLP) has been characterized by the development of \textit{larger} language models \cite{bender2021dangers} requiring substantial computational resources and access to extensive datasets that have been challenging to obtain outside big companies and well-funded research labs. As a result, large language models are increasingly organized within a supply chain and produced by several interconnected actors. Cobbe et al. \cite{cobbe2023understanding} describe the algorithmic supply chain as a pipeline where  \textit{`several actors contribute towards the production, deployment, use, and functionality of AI technologies.'} In the case of large language models, companies providing LLMs `as a service’ offer access to pre-built general-purpose models. As a result, LLM providers hold a lot of control over the distribution and development of the underlying technologies, including how to identify and address ethical concerns. Researchers in our study, situated downstream within the algorithmic supply chain, described a lack of control to examine and mitigate ethical issues. While some expressed a lack of desire, others struggled to pay off ethical debt \cite{petrozzino2021pays} accrued through components developed upstream in the supply chain. How might we then interrupt the supply chain and shift control downstream: could we imagine tools where researchers can set up and maintain the infrastructures needed to integrate large language models in their workflow? How might such tools center a proactive approach to identify ethical concerns?

We recognize the barriers in navigating ethical concerns when researchers depend on AI service providers for access to these tools. There is a significant practical cost associated with choosing a language model that attends to ethical concerns. Our findings suggest opportunities to create tools, methods and processes to support HCI researchers in better navigating the ethical considerations of integrating language models into their projects. For example,  known evaluation frameworks for the use of LLMs in social science research such as CoMPosT~\citep{cheng-etal-2023-compost} make LLMs’ limitations transparent to researchers, aiding them in regaining control over the LLM tools. In addition to validation tools for LLMs for research-specific purposes, there is a growing call for auditing LLMs across technology providers, model development, and downstream applications, covering the LLM supply chain ecosystem~\citep{M_kander_2023}. Auditing tools of LLMs are crucial for overseeing whether technology providers maintain ethical standards and ensure responsible development of LLMs including transparent policies on user data handling.

A potential issue of depending on the black-box LLMs from service providers, which is also discovered in our findings, is the risk of inheriting the biases and misinformation from upstream in the LLM supply chain. To address these ethical challenges, it is essential to develop frameworks and tools that empower researchers to host and manage their own LLMs for research. There has been research providing a comprehensive and practical guide for practitioners and end-users to work with LLMs~\cite{yang2023harnessing}, with a focus on closed-source LLMs such as ChatGPT. In order to offer more autonomy and control for researchers in integrating LLMs into their research workflow, new technological support in streaming that process is needed. By providing these capabilities, researchers would be able to actively engage with and mitigate the ethical concerns. 

\textbf{Creating learning opportunities and materials for ethics of LLM use in HCI. } Researchers' awareness of ethical issues related to LLMs influenced how they address these considerations in their projects. Many researchers spoke of LLMs as an unfamiliar territory, lacking the training and well-established guidelines to approach ethical considerations with this emerging technology. We could draw on lessons from scholarship in FAccT, STS, NLP and other allied domains, as well as industry resources, to develop learning opportunities, such as toolkits \cite{shen2021value}, guidebooks \cite{PAIRGuidebook}, or frameworks \cite{derczynski2023assessing, ai2023artificial}, for understanding and addressing ethical concerns with using LLMs. Achieving this requires particular attention from the HCI community to collaboratively explore how LLMs are used in different stages of HCI research. 

Conferences are a valuable starting point to promote cross-institutional learning. We propose organizing workshops and panels (such as \cite{shen2023shaping}) with interdisciplinary experts in sociotechnical understanding of LLMs to raise awareness on the impacts of using LLMs. Additionally, we call for creating and disseminating case studies (\textit{e.g.,} \cite{PAIRcasestudies}), potentially included in newsletters, illustrating how HCI researchers actively address ethical concerns related to LLMs. A repository of case studies, covering diverse HCI domains and epistemologies, might also offer resources for preventing and mitigating ethical concerns. Education on research ethics for LLM use should be a necessary component of HCI research and practice, given the well-documented risks and adverse impacts of LLMs \cite{taxonomy-risks}. One avenue is incorporating such case studies into HCI curriculum, especially if targeted towards research students, to provide real-world examples that help build their awareness of these issues.

\textbf{Shifting academic incentives to foreground ethical concerns. } 
Throughout the interviews, we observed that researchers could generally foresee potential ethical concerns with LLMs, often drawing from their familiarity with the literature or discussions with other researchers. Nonetheless, some participants went on to articulate their reason for not prioritizing these ethical considerations in their projects. Researchers described how constraints such as limited funding, pressure to publish, conference deadlines, often came in the way of focusing on ethical concerns. The need to address ethical issues could then be relegated to the limitations or future work section, reflecting broader perceptions within Computer Science research that ethics are often viewed as secondary considerations. Indeed, Do and Pang \textit{et al.} \cite{do2023s} discussed how academics may have lower incentive to examine unintended consequences of their research (also observed in our study) in contrast with industry practices.

There is a renewed urgency to reconsider research ethics practices within HCI and how we navigate challenges presented by emerging technologies such as LLMs. We take inspiration from Do and Pang \textit{et al.} \cite{do2023s} and Soergel \textit{et al.} \cite{soergel2013open} to propose starting points for changing structural incentives within academia. Publications and citations are the currency and means for upward mobility in research. Could we shift criteria for recognition and funding to explicitly recognize attention to ethical considerations in research practices? Publishing organizations, funding agencies, and regulatory institutions will pay a crucial role in reshaping incentives. Most importantly, HCI researchers must acknowledge the ways in which they actively shape the discourse around associated risks and ethical considerations through their research, and work actively towards a cultural shift that demonstrates a commitment to ethical use of LLMs.    

\section{Limitations and Future Work}

Our study sheds light on the emerging practices regarding LLM ethics among HCI researchers, but it has limitations due to its exploratory nature. Firstly, while we sought to offer a deep understanding on how HCI researchers navigate LLM ethics, our sample was constrained by our snowball sampling method. HCI research encompasses a wide range of diverse research traditions, many of which we were unable to include in our study. This limitation highlights the need for more comprehensive and systematic future studies in this area. Our interview sample primarily consists of researchers from the USA, and future research may want to further explore the ethical challenges and practices related to the use of LLMs by researchers from other regions. Secondly, there's a potential selection bias in our study: we may have primarily attracted respondents who are conscious of their LLM usage and are open to discussing their experiences in a research setting. To gain a broader perspective, future research should explore how ethical practices with LLMs vary across different research methodologies, domains, and settings, including both industry and academia. 
\section{Conclusion}
In this paper, we drew empirical data from a survey and interviews to explore how HCI researchers have currently integrated LLMs into their research practices, what ethical concerns they have encountered as well as how they've navigated those concerns. Our results suggested that although HCI researchers have used LLMs across their research processes and are aware of a wide variety of ethical concerns, in many cases, they have challenges in effectively identifying and navigating those concerns in their own projects. Reflecting on these findings, we discuss potential approaches to support the formation of emerging ethical norms for using LLMs in HCI research. We encourage HCI researchers to proactively engage with IRB and collaborate with policymakers and generative AI companies on creating guidelines for the responsible use of LLMs. We also identify the need to re-examine the informed consent process and provide technological support to interrupt the LLM supply chain. In addition, we discuss the importance of creating learning opportunities for the ethics of LLMs use in HCI and shifting academic incentives to prioritize ethical concerns.

\bibliographystyle{ACM-Reference-Format}
\bibliography{references}


\end{document}